\documentclass[aps,prd,twocolumn,amsmath,amssymb,amsfonts,nofootinbib,superscriptaddress,altaffilletter,showpacs,floatfix]{revtex4-1}
\usepackage{amsgen,amstext,amsbsy,amsopn,amsfonts,amssymb}
\usepackage{supertabular}
\usepackage{graphicx}
\usepackage{hyperref}
\hypersetup{bookmarksopen=true}
\usepackage{graphpap}
\usepackage{rotating}
\usepackage{url}
\usepackage{color}
\usepackage{dcolumn}
\def\UL{\textrm{UL}}

\def\etal{{\it et al.}}

\def\sci#1#2{{#1}{\times 10^{#2}}}

\def\hx{h_{\times}}
\def\hp{h_+}

\def\Ap{A_+}
\def\Ax{A_{\times}}

\def\Fp{F_+}
\def\Fx{F_{\times}}

\def\Re{\textrm{Re}\,}
\def\Im{\textrm{Im}\,}

\def\UL{\textrm{UL}}
\def\ULh{\widehat{\UL}}
\def\LL{\textrm{LL}}
\def\LLh{\widehat{\LL}}

\def\fpp{f_{pp}}
\def\fpc{f_{pc}}
\def\fcc{f_{cc}}
\def\fipc{f_{impc}}

\definecolor{green4}{rgb}{0,0.25,0}

\def\dochecks{0}
\def\check#1{{\if \dochecks 1 {\color{green4}\ \\ \rule{0.4\linewidth}{1pt}\hfill Check \hfill\rule{0.4\linewidth}{1pt} \hspace*{0pt} \par #1 \rule{\linewidth}{1pt} } \fi}}

\begin{document}
\title{Efficient representations of upper limits and confidence intervals for experimental data}
\author{Vladimir Dergachev}
\email{vladimir.dergachev@aei.mpg.de}
\affiliation{Max Planck Institute for Gravitational Physics (Albert Einstein Institute), Callinstrasse 38, 30167 Hannover, Germany}
\affiliation{Leibniz Universit\"at Hannover, D-30167 Hannover, Germany}

\begin{abstract}
Upper limits and confidence intervals are a convenient way to present experimental results. With modern experiments producing more and more data, it is often necessary to reduce the volume of the results for convenient distribution. A common approach is to take a maximum over a set of upper limits, which yields an upper limit valid for the entire set. This, however, can result in significant sensitivity loss, because we discard more constraining data. In this paper we introduce {\em functional} upper limits and confidence intervals that allow to summarize results with small relaxation of constraints. A toy example and an application to upper limits in all-sky continuous gravitational wave searches are worked out.

\end{abstract}
\maketitle

\section{Introduction}
Many modern experiments and surveys produce large amounts of data, whose dimensionality keeps increasing. For example, Gaia \cite{gaia_dr3} and LIGO \cite{aligo} collect terabytes of data which is postprocessed and reduced in size for public release. A particular challenge was encountered during construction of the atlas of continuous gravitational waves \cite{o3a_atlas2a}. This atlas is produced from the the data of an all-sky search that computes billions of separate polarization-dependent upper limit curves. These curves are impractical to store in raw form and were not included in the first atlas release \cite{o3a_atlas1}.

One way to reduce the storage requirements is to employ data compression \cite{shannon, lz, jpeg2000, dct}. However, lossless compression is often only moderately effective, while conventional lossy compression can make upper limits (and confidence intervals in general) invalid. 

What is needed is a way to perform lossy compression while retaining the validity of upper limits and confidence intervals. This is clearly possible by computing maxima and minima of upper and lower bounds; however, this can result in significant sensitivity loss, because we discard more constraining data.

In this paper, we introduce the notion of {\em functional} upper limits, which replace the overall maximum with a majorant - a function uniformly above original upper limit values - that can be specified with a few numbers.

We describe a general method to construct such {functional} upper limits and a straightforward generalization to lower limits and confidence intervals. A toy example is provided to illustrate the method. 

This is followed by a fully worked out study of functional upper limits on continuous gravitational wave radiation. This paper also serves as a reference description of functional upper limit data for releases of atlas of continuous gravitational waves \cite{o3a_atlas2a}. We comment on technical issues overcome during production of billions of functional upper limit curves.

\section{Functional compression}

Suppose our search has computed upper limits $\UL=\UL(x)$ as a function of variable $x \in X$. 

To compress the data, we pick a set of functions $\left\{\ULh\left(x ; \vec{c}\right)\right\}$ depending on coefficients $\vec{c}$. The simplest example of such a set is a finite-dimensional linear space, but it is useful to consider more complicated spaces such as an image of a linear space under a differentiable function.

The goal of our compression algorithm is to find a parameter $\vec{c}$ such that the compressed form is always at or above original upper limit data:
\begin{equation}
\label{eqn:validity}
\forall x \in X : \ULh \left(x ; \vec{c}\right)\ge \UL \left(x\right)
\end{equation}
while minimizing error
\begin{equation}
\label{eqn:error1}
\max_{x \in X} \left( \ULh \left(x ; \vec{c}\right) - \UL \left(x\right)  \right) \rightarrow \min
\end{equation}
The choice of error function is somewhat arbitrary. It is often a good idea to pick an error function that makes the optimization problem easy to compute. This, of course, depends on the optimization algorithm.

The compression of lower limits is done in the same way, by simply inverting the sign. Confidence intervals can be compressed by treating lower and upper bounds separately.

There is an efficient way to compute optimal function $\ULh$. 
To begin, suppose we pick a set of constant functions $\left\{ c : c \in {\mathbb R} \right\}$. Then the compression algorithm can compute
\begin{equation}
\ULh = c = \max_{x} \UL\left(x\right) 
\end{equation}
This clearly satisfies the validity constraint given by Eqn. \ref{eqn:validity} and is easy to compute. However, the error equals to the range of upper limit values.  

In situations where some parameters $x$ produce smaller upper limits than others, we can compress the range by introducing normalization $g(x)$:
\begin{equation}
c=\max_{x} \left(\UL\left(x\right) g(x) \right)
\end{equation}
Now the one-parameter set of functions is $\left\{c/g(x) : c \in {\mathbb R}\right\}$.

We can further improve the performance by introducing more coefficients:
\begin{equation}
\ULh(x) = h\left(\frac{\vec{c} \cdot \vec{f}\left(x\right)}{g(x)}\right) 
\end{equation}
where $h$ is an increasing monotonic invertible function and $\vec{f}$ and $g$ are some suitable functions of $x$. At this point one could assume $g(x)=1$ and just redefine $\vec{f}(x)$, but, in practice, deciding on suitable functions $\vec{f}$ is a lot easier with normalization first taken out. In the fully worked out examples below we use $f_1(x)=1$, but one could envision problems where a different choice reduces estimation error, once $g(x)$ is fixed.

With these definitions, the problem of finding optimal coefficients $\vec{c}$ can be framed as a linear programming problem:
\begin{equation}
\left\{
\begin{array}{l}
h^{-1}\left(\UL(x)\right) g(x) \le \vec{c} \cdot \vec{f}(x) \\
\underset{x\in X}{\max}\left(\frac{\vec{c} \cdot \vec{f}\left(x\right)}{g(x)}-h^{-1}\left(\UL(x)\right)\right)\rightarrow \min
\end{array}\right.
\end{equation}
as long as we are willing to modify the error function (eqn. \ref{eqn:error1}) to compute the error in $h^{-1}\left(\UL\right)$.

Theoretically, this data compression can be very effective because we replace a large vector of values $\left(\UL(x): x\in X\right)$ with much smaller number of coefficients $\vec{c}$. And since coefficients $\vec{c}$ can be found with a linear programming algorithm, this technique is applicable to large volumes of data, with functional upper limits computed repeatedly for different inputs $\UL(x)$.

The use of more advanced solvers, for example capable of solving quadratic or mixed-integer problems would provide the freedom to state constraints on coefficients $\vec{c}$ or the optimality condition in a different way. These algorithms often have large computational costs which need to be balanced against any improvements in data compression. One possibility is to first compress the data with a fast solver, assess compression efficiency and only employ more sophisticated (and slower) solvers on the subset of data where the gains in compression are worthwhile.

The choice of functions $\vec{f}(x)$, $g(x)$ and $h(v)$ is highly specific to the data being compressed. 
Mathematically, for a given number of coefficients $\vec{c}$ there should be an optimal set of functions $\vec{f}(x)$ that minimizes the error. However, finding them analytically is only possible in a few idealized situations. 

The author's approach is to first start with functions of some obvious significance. 
For example, a constant function can describe the common level and a linear one can describe a trend or a moment. If physically inspired functions are not enough to get 
small overestimates, a good idea is to add some suitable harmonics. These could be Fourier or spherical harmonics for linear or spherical parameter spaces, Chebyshev functions for bounded segments, or others, depending on the parameter space and optimization problem. 

Once enough functions have been found to get overestimate under control, one can start optimizing the number of parameters $\vec{c}$ by excluding functions a few at a time to judge their usefulness.

\section{Toy example}

Suppose, in some experiment, we are interested in measuring a quantity $y=y(x)$ depending on parameter $x$ that varies continuously between $0$ and $1$. The measurement is carried out by constructing a grid of $N+1$ points $\left\{x_k=k/N: k=0 \ldots N\right\}$ and then establishing upper limits $\UL(x)$ and lower limits $\LL(x)$ on $y(x_k)$ individually for every $k$.

We would like to construct functional upper and lower limits for $y(x)$. Since this is 
a generic example we do not have any domain-specific information on $y(x)$. Therefore, we choose $h(y)=y$ and $g(x)=1$.

Similarly, in the absence of any additional information on $y(x)$ we have a lot of freedom to choose $\vec{f}(x)$. Some good choices would be trigonometric functions or orthogonal polynomials. As trigonometric functions are often expensive to compute, let us choose orthogonal polynomials of degree at most $M$. This choice is equivalent to choosing $f_m(x)=x^m$ for $m=0 \ldots M$. 

Therefore, in order to compress upper limits we need to solve the following optimization problem:

\begin{equation}
\left\{
\begin{array}{l}
\UL(x_k) \le  \sum_{m=0}^{M} c_m^\UL f_m(x_k) \\
\underset{k=0 \ldots N}{\max}\left(\sum_{m=0}^{M} c_m^\UL f_m(x_k)-\UL(x_k)\right)\rightarrow \min
\end{array}\right.
\end{equation}
Here we have $N+1$ inequalities, one for every $x_k$.

The $\max \rightarrow \min$ form is not quite suitable as input to numerical linear solvers. To fix this we introduce an auxiliary variable $u$ and rewrite our optimization problem as follows:

\begin{equation}
\left\{
\begin{array}{l}
\UL(x_k) \le  \sum_{m=0}^{M} c_m^\UL f_m(x_k) \\
\sum_{m=0}^{M} c_m^\UL f_m(x_k)-\UL(x_k) \le u \\
u\rightarrow \min
\end{array}\right.
\end{equation}
This is an optimization problem with $M+2$ variables and $2N+2$ constraints, that can be readily passed to a numerical solver. The problem is  always feasible, as choosing $c_m=0$ for $m>0$ and $c_0=u=\max_k \UL(x_k)$ gives a valid but not necessarily optimal solution.

For lower limits we solve the optimization problem:
\begin{equation}
\left\{
\begin{array}{l}
\LL(x_k) \ge  \sum_{m=0}^{M} c_m^\LL f_m(x_k) \\
\LL(x_k)-\sum_{m=0}^{M} c_m^\LL f_m(x_k) \le u \\
u\rightarrow \min
\end{array}\right.
\end{equation}

The coefficients $c_m^\UL$ and $c_m^\LL$ are sufficient to compress gathered data for upper and lower limits in points $x_k$. However, what we would really like is to have functional limits valid in every point $x$, even outside the grid. 

For this we need some information on how $y(x)$ behaves for nearby points. A common situation is when $y(x)$ is Lipschitz with constant $L$:
\begin{equation}
|y(x)-y(x_k)|\le L |x-x_k| 
\end{equation}
This is true for any differentiable $y(x)$ which derivative is bounded $|y'(x)|\le L$. 
However, the Lipschitz condition is wider than just differentiable functions, for example $|x|$ is Lipschitz. 

In situations, when $y(x)$ could have small jumps, or contains some stochastic component, we can consider a generalization of Lipschitz condition with parameter $\delta$:
\begin{equation}
|y(x)-y(x_k)|\le L |x-x_k|+\delta
\end{equation}
For $\delta=0$ we get our original Lipschitz condition.

Assuming $y(x)$ satisfies the condition above we can now construct functional upper and lower limits:
\begin{equation}
\begin{array}{l}
\ULh(x)=\sum_{m=0}^M c_m^\UL f_m(x)  + \frac{L}{2N}+\delta \\
\LLh(x)=\sum_{m=0}^M c_m^\LL f_m(x)  - \frac{L}{2N}-\delta \\
\end{array}
\end{equation}
These upper and lower limits are valid for any $x\in[0,1]$.

In the following sections we will describe the more complicated construction of the functional upper limits distributed with the atlas of continuous gravitational waves.

\section{All-sky searches for continuous gravitational waves}

All-sky searches for continuous gravitational waves produce vast amounts of data \cite{o3a_atlas1, o3a_atlas2a, O2_falcon, O2_falcon2, O2_falcon3, keith_review, EatHO3a, lvc_O3_allsky, lvc_O3_allsky2}. These searches are looking for unknown gravitational wave sources by sweeping large parameter spaces. Of course, most of the parameter space does not contain signals loud enough to be detected. In this case, the search places an upper limit on possible signal strength.

Reporting these upper limits can be problematic because of large number of dimensions - a relatively simple search will have two dimensions for the sky, one for frequency and two for polarization. Allowing additional parameters, such as frequency derivative or binary evolution of the source, increases dimensionality further. 

The technique used in the past to compress the data was to maximize upper limits over a subset of parameter space. For example, a maximum can be computed over polarization and a small range of frequencies.

\begin{figure}[htbp]
\centering
\includegraphics[width=3.3in]{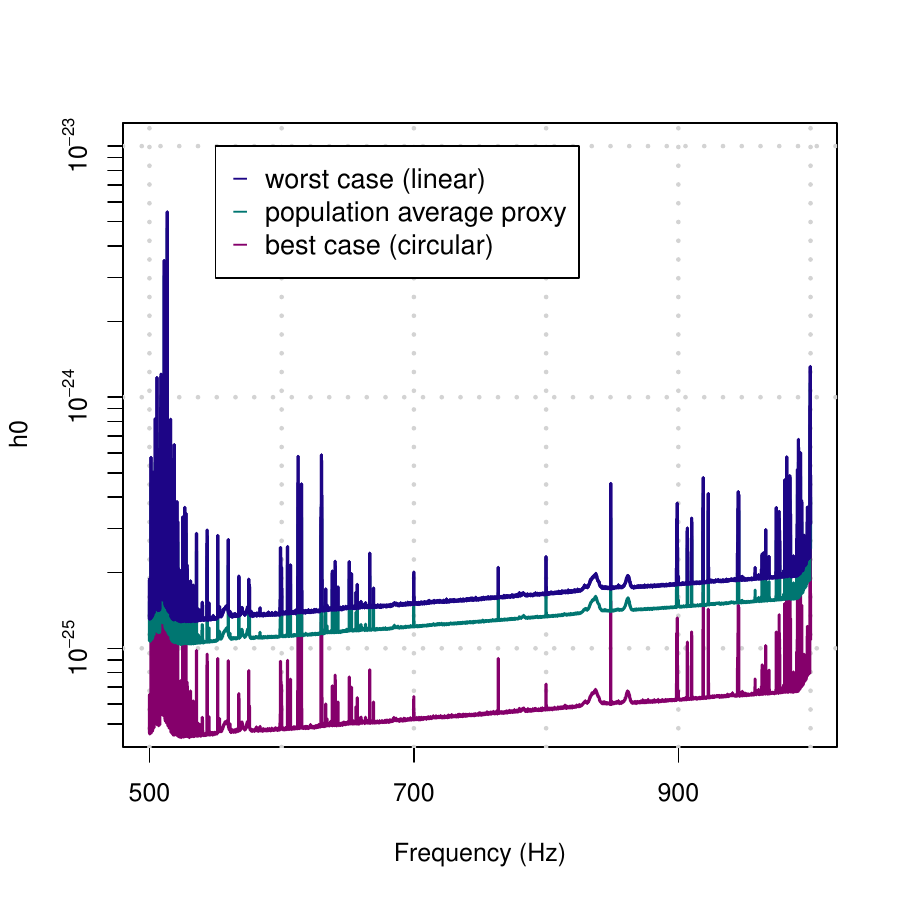}
\caption[Upper limits]{
\label{fig:O3_atlas1}
Gravitational wave intrinsic amplitude $h_0$ upper limits at 95\% confidence as a function of signal frequency \cite{o3a_atlas1}.
}
\end{figure}

Reducing polarization data by maximization is far from optimal because there is a large variation in upper limits between linearly and circularly polarized sources, as can be seen in Figure \ref{fig:O3_atlas1} which shows result of maximization over sky, frequency derivative, small frequency band and polarization. The three curves in the figure correspond to different choices in treating polarizations - the upper curve gives worst case upper limit derived from maximization over all polarizations, the middle curve is polarization average proxy which approximates upper limit on a source with random polarization and random sky coordinates, and the bottom curve shows maximum over two circular polarizations.

%In this paper we describe how one can report polarization specific upper limits using only a few dozen numbers per record. These numbers define a function of coefficients computed from polarization parameters.

\section{Polarizations of gravitational waves}

A prototypical source of continuous gravitational waves is a neutron star with equatorial deformation. Such rotating source will emit circularly polarized waves along the rotation axis and linearly polarized wave, perpendicular to the rotation axis.

The continuous gravitational wave upper limits are stated as upper limits on the amplitude of such waves. In this section we introduce complex amplitude parameters $\tilde{w}_{1,2}$, algebraic combinations of which form functions $\vec{f}$.

We start by assuming that our signal consists of two polarizations:
\begin{equation}
\begin{array}{rcl}
\hp'&=&\Ap\cos(\omega t+\phi) \\
\hx'&=&\Ax\sin(\omega t+\phi) \\
\end{array}
\end{equation}

A generic pulsar signal can be represented as $\Ap=h_0\left(1+\cos^2(\iota)\right)/2$, 
$\Ax=h_0\cos(\iota)$, with $h_0=\Ap+\sqrt{\Ap^2-\Ax^2}$ and 
$\cos(\iota)=\Ax/\left(\Ap+\sqrt{\Ap^2-\Ax^2}\right)$.

We will assume that demodulation is performed for a fixed frame of plus and cross polarizations rotated at an angle $\beta$. 
In this coordinate system we have:
\begin{equation}
\begin{array}{rcl}
\hp&=&\Ap\cos(\omega t+\phi)\cos(\epsilon)-\Ax\sin(\omega t+\phi)\sin(\epsilon) \\
\hx&=&\Ap\cos(\omega t+\phi)\sin(\epsilon)+\Ax\sin(\omega t+\phi)\cos(\epsilon) \\
\end{array}
\end{equation}
where $\epsilon=2(\psi-\beta)$. Here $\phi$ describes initial signal phase. The angle $\psi$ describes polar angle of the source rotation axis to the line of sight. The angle $\beta$ is important when performing analysis because the detector orientation changes with time. For reporting results, we can simply pick any convenient frame, so for the rest of the paper we assume $\beta=0$. 

Continuous wave searches usually do not use high sample rate data. Rather, a set of short Fourier transforms (SFTs) is computed, and the search loads only the bins covering the frequency band of interest. The Fourier transforms are picked to be short enough so that the change in Doppler modulation from Earth motion over the duration of the SFT is negligible.

An interferometric gravitational wave detector \cite{aligo} measures differences in length change between two fixed perpendicular directions. Its response to the signal is described by two functions $\Fx(t)$ and $\Fp(t)$. These functions are slowly varying and can be assumed to be constant for the duration of one SFT. 

The signal amplitude in SFT bin corresponding to frequency $\omega$ can now be computed as
\begin{equation}
\begin{array}{rcl}
z&=&\displaystyle\int \left(\Fp\hp+\Fx\hx\right) e^{-i\omega t} dt=\\
 &=&\vphantom{\int}\frac{1}{2}e^{i\phi}\left(\Fp(\Ap \cos(\epsilon)+i\Ax \sin(\epsilon))+\right.\\
 &&\qquad\left.+\Fx(\Ap \sin(\epsilon)-i\Ax \cos(\epsilon))\right)\\
&=&\Fp w_1+\Fx w_2
\end{array}
\end{equation}
where we have introduced complex amplitude parameters
\begin{equation}
\begin{array}{rcl}
w_1&=&\frac{1}{2}e^{i\phi}(\Ap \cos(\epsilon)+i\Ax \sin(\epsilon))\\
w_2&=&\frac{1}{2}e^{i\phi}(\Ap \sin(\epsilon)-i\Ax \cos(\epsilon))\\
\end{array}
\end{equation}
The complex amplitude parameters $w_1$ and $w_2$ are algebraically symmetric, and satisfy the following equation of constant $h_0$:

\begin{equation}
\sqrt{|w_1+iw_2|}+\sqrt{|w_1-iw_2|}=\sqrt{h_0}
\end{equation}
the solutions of which form a singular surface enclosing a non-convex solid. This complicated shape is responsible for differences between worst-case upper limits, population average upper limits, and circularly polarized upper limits that are seen in Figure \ref{fig:O3_atlas1}.

The normalized complex parameters are introduced by setting $\tilde w_1=w_1/h_0$ and $\tilde w_2=w_2/h_0$, and satisfy the equation:
\begin{equation}
\sqrt{|\tilde w_1+i \tilde w_2|}+\sqrt{|\tilde w_1-i \tilde w_2|}=1 
\end{equation}
We will use the normalized complex parameters to construct the function space of majorants in the following section.

\section{Construction of functional upper limits}
The polarization dependence of upper limits is described by a function on two-dimensional space $x=(\iota, \psi)$.

We pick $h(v)=\sqrt{v}$, thus we measure error in the square of the upper limit value. 
The vector $\vec{c}$ has 14 components $c_{1-14}$.
Our upper limits model $\ULh$ is then a square root of a rational function of trigonometric functions of $\iota$ and $\psi$:
\begin{equation}
\begin{array}{l}
\label{eqn:model}
\ULh^2 =  \left(c_1+\fpp c_2+\fpc c_3+\fcc c_4+\fipc c_5+\fpp ^2c_6+\right.\\
\quad\quad\quad\left.+\fcc ^2c_7+\fpc ^2c_8+\fipc \fpp c_9+\fipc \fpc c_{10}+\right.\\
\quad\quad\quad\left.+\fipc \fcc c_{11}+\fpp \fpc c_{12}+\fcc \fpc c_{13}+\right.\\
\quad\quad\quad\left.+\fpp \fcc c_{14}\right)/\left(\fpp +\fcc \right)
\end{array}
\end{equation}
where coefficients $f_{\cdot}=f_{\cdot}(\iota, \psi)$ are given by the following formulas:
\begin{equation}
\begin{array}{ll}
 a_{+} & = \frac{\left(1+\cos^2 \iota\right)^2}{4}\\
 a_{\times} & = \cos^2 \iota \\
 \fpp  &= 2 |\tilde w_1|^2=\frac{1}{4}\left(a_+ + a_\times + \left(a_+-a_\times\right)\cos 4\psi\right)\\
 \fpc  &= 4\Re (\tilde w_1 \tilde w_2^*)=\frac{1}{2}\left(\left(a_+-a_\times\right)\sin 4\psi\right) \\
 \fcc  &= 2 |\tilde w_2|^2=\frac{1}{4}\left(a_+ + a_\times - \left(a_+-a_\times\right)\cos 4\psi\right) \\
 \fipc  &= 2\Im (\tilde w_1 \tilde w_2^*)= \frac{1}{4}\left(1+\cos^2 \iota\right)\cos \iota
\end{array}
\end{equation}
The denominator of equation \ref{eqn:model} helps to compress the range of upper limits, while coefficients $c_{1-14}$ allow sufficient freedom to bring upper limit overestimate under 5\% for most data.

The coefficients $c_{1-14}$ can be found using the following linear optimization problem:
\begin{equation}\label{linmodel}
\left\{
\begin{array}{l}
\UL^2\left(\fpp +\fcc\right)\le c_1+\fpp c_2+\fpc c_3+\fcc c_4+\\
\quad\quad\quad+\fipc c_5+\fpp ^2c_6+\fcc ^2c_7+\fpc ^2c_8+\\
\quad\quad\quad+\fipc \fpp c_9+\fipc \fpc c_{10}+\fipc \fcc c_{11}+\\
\quad\quad\quad+\fpp \fpc c_{12}+\fcc \fpc c_{13}+\fpp \fcc c_{14}\\
\UL^2\left(\fpp +\fcc\right)\ge c_1+\fpp c_2+\fpc c_3+\fcc c_4+\\
\quad\quad\quad+\fipc c_5+\fpp ^2c_6+\fcc ^2c_7+\fpc ^2c_8+\\
\quad\quad\quad+\fipc \fpp c_9+\fipc \fpc c_{10}+\fipc \fcc c_{11}+\\
\quad\quad\quad+\fpp \fpc c_{12}+\fcc \fpc c_{13}+\fpp \fcc c_{14}+\\
\quad\quad\quad+u w \\
u \rightarrow \min
\end{array}\right.
\end{equation}
here $\UL$, $f_{\cdot}$ and $w$ are functions of $\iota$ and $\psi$, typically taken from precomputed grid. The auxiliary variable $u$ has been introduced with weight $w$ to control error. After optimization, a correction can be applied that compensates for grid spacing and assures that upper limits $\tilde\UL$ are valid for any $\iota$ and $\psi$. The simplest way to compute the correction is to bound the worst-case power loss from projection of the signal onto nearby grid template. This behaves as a cosine of the mismatch angle and rapidly converges to unity as the grid spacing is decreased. %In other words, the square of the upper limit is Lipschitz with $L=1$.

There are many existing linear programming algorithms  \cite{bellman, glpk, ortools} capable of solving model \ref{linmodel}, in particular, various simplex algorithms and interior point methods. As the model has few variables, it appears particularly simple.

One implementation quirk that makes things interesting is that practical implementations of linear programming algorithms often encounter difficulties on some subset of inputs. For example, simplex algorithms can loop and/or fail to converge due to ill-conditioned matrices that define the problem. 

In many situations, this is not a big difficulty as the input data can be slightly perturbed. However, the all-sky searches such as Falcon \cite{o3a_atlas1} need to 
find coefficients $c_{1-14}$ for billions of models, which input data, $\UL$, is derived from noise. 

This practically guarantees that any trouble spot of the chosen linear optimization algorithm will be encountered during the analysis. The solution we have used is to try different optimization algorithms \cite{glpk, ortools} one after another with a time limit within which they are expected to converge. Finally, in rare situations when none of these succeed we settle for a solution of \ref{linmodel} that is valid, but not necessarily optimal.

\section{Application example}

To test our technique, we implemented it as part of a Falcon search and then carried out a search on data with software injections. The 31574 injections were uniformly distributed in the sky, with frequency derivatives from 0 to $\sci{-5}{-10}$\,Hz/s, and polarizations distributed assuming random source orientations. 

\begin{figure*}[htbp]
\centering
\includegraphics[width=7in]{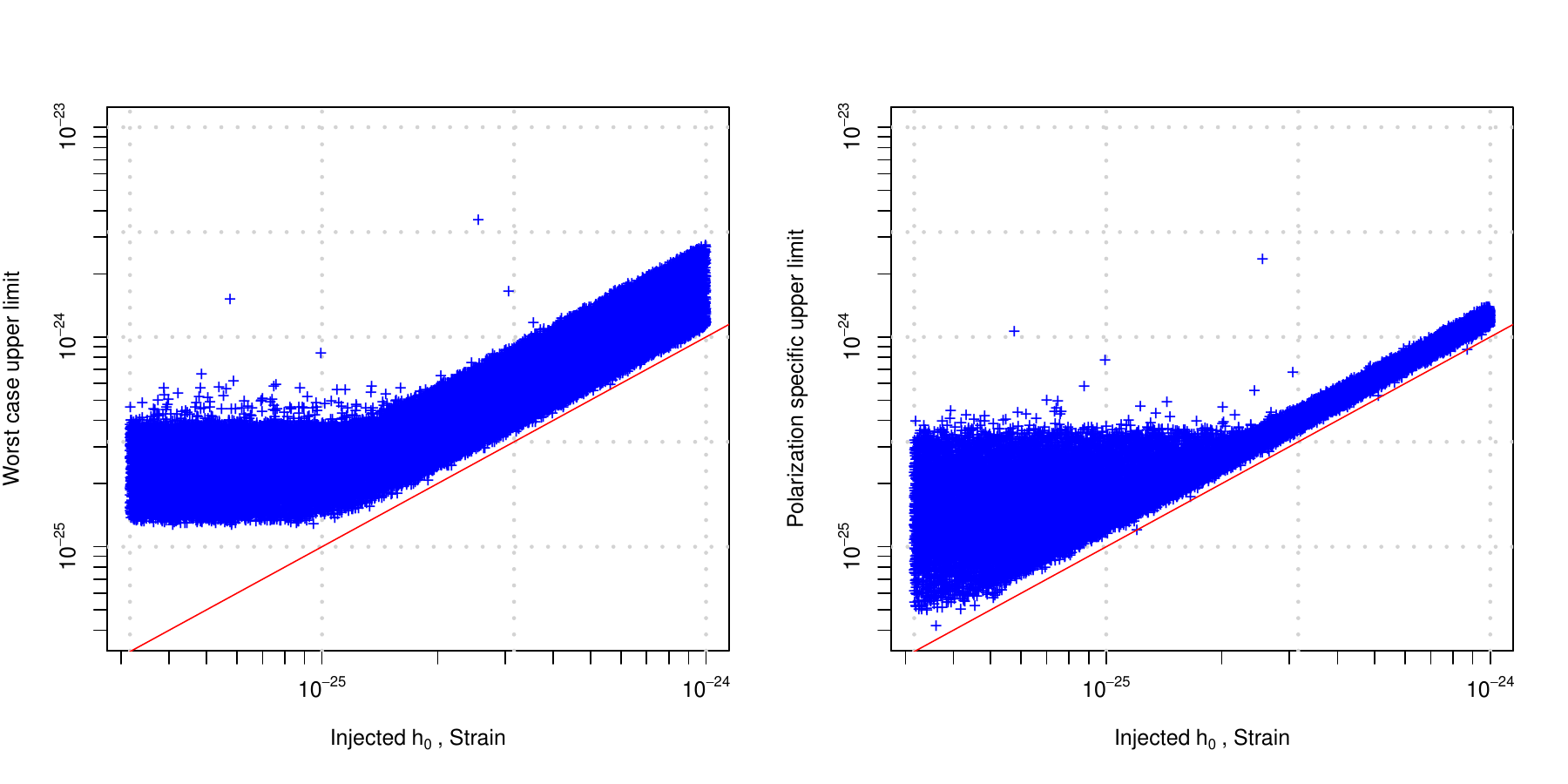}
\caption[Upper limits]{
\label{fig:upper_limit_recovery}
Upper limit versus injected strain for a set of software injections \cite{o3a_atlas2a}. The plot on the left shows worst case upper limit, while the graph on the right shows polarization specific upper limit computed using $\iota$ and $\psi$ used to inject signals. The red line shows injected strain values. The plot on the right has appeared previously in \cite{o3a_atlas2a}.
}
\end{figure*}

Figure \ref{fig:upper_limit_recovery} shows how upper limits established by Falcon pipeline compare with injection strength. To be correct, the upper limit has to be above the $y=x$ line marked in red. 

\begin{figure}[htbp]
\centering
\includegraphics[width=3.3in]{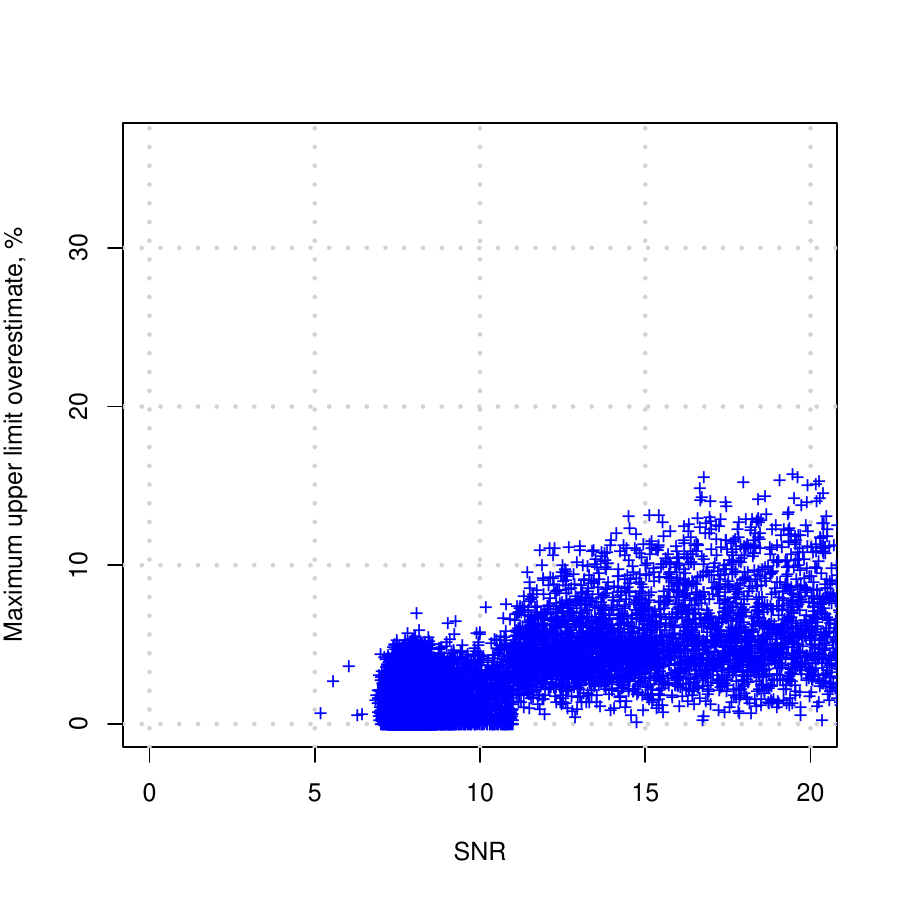}
\caption[Upper limits]{
\label{fig:rel_err}
Estimate of relative upper limit error versus SNR of software injection. For injections with small SNR the worst case error is around 5\%, for stronger injections it can be much larger.
}
\end{figure}

The left plot shows conventional worst-case upper limit. We see a fairly wide spread above the red line, which happens because some injections are circularly polarized and deliver a lot more power. 

The right plot shows polarization specific upper limit. To produce it, the Falcon pipeline first compressed the data to produce a functional upper limit, and then that function was evaluated for $\iota$ and $\psi$ that were used to make the injection.
We see that polarization specific upper limits have a far smaller spread for loud injections as they are more accurate.

For very weak injections the injected signals are dominated by detector noise. Here, polarization specific upper limits show their main benefit as they can rigorously establish far lower upper limits for elliptically polarized signals. Indeed, the spread of blue points extends far below $10^{-25}$.

Figure \ref{fig:rel_err} shows how relative error in the upper limit depends on the injection strength. The functional model has been optimized for the noise dominated case. There we overestimate upper limits by 5\% or less. As injection signal-to-noise ratio (SNR) grows, our model becomes less efficient. This is not a big problem because for SNRs that large the Falcon pipeline can easily recover astrophysical signals.

\section{Conclusions}
We have described a new method that allows to establish upper limits and confidence intervals as a function, rather than a single number. This method has been successfully used to compress Falcon atlas of continuous gravitational waves, where for the first time, we were able to provide upper limits for {\em all} polarizations. The functional upper limits are established by linear optimization that, in rare cases, presents difficulties for conventional linear optimization codes. It would be interesting to explore whether a more reliable optimizer can be constructed to solve the model of the form \ref{linmodel}.

\end{document}